\documentclass[letterpaper]{article} 
\usepackage{aaai25}  
\usepackage{times}  
\usepackage{helvet}  
\usepackage{courier}  
\usepackage[hyphens]{url}  
\usepackage{graphicx} 
\usepackage{url}
\usepackage[hang,flushmargin]{footmisc} 
\usepackage{natbib}  
\usepackage{caption} 
\usepackage{algorithm}
\usepackage{algorithmic}

\usepackage{newfloat}
\usepackage{listings}
\DeclareCaptionStyle{ruled}{labelfont=normalfont,labelsep=colon,strut=off} 
\floatstyle{ruled}
\newfloat{listing}{tb}{lst}{}
\floatname{listing}{Listing}
\usepackage{xcolor}
\newcommand{\answerYes}[1]{\textcolor{blue}{#1}} 
\newcommand{\answerNo}[1]{\textcolor{teal}{#1}} 
\newcommand{\answerNA}[1]{\textcolor{gray}{#1}} 
 
\title{How Influencers and Multipliers Drive Polarization and Issue Alignment on Twitter/X\\[1mm]}
\author {
    Armin Pournaki\textsuperscript{\rm 1,2,3},
    Felix Gaisbauer\textsuperscript{\rm 4},
    Eckehard Olbrich\textsuperscript{\rm 1}
}
\affiliations {
    \vspace{2mm}
    \textsuperscript{\rm 1}Max Planck Institute for Mathematics in the Sciences, Leipzig, Germany\\
    \textsuperscript{\rm 2}Laboratoire Lattice, École Normale Supérieure - PSL - CNRS - Univ. Sorbonne Nouvelle, Montrouge, France\\
    \textsuperscript{\rm 3}médialab, Sciences Po, Paris, France\\
    \textsuperscript{\rm 4}Weizenbaum Institute for the Networked Society, Berlin, Germany\\
    \vspace{2mm}
    pournaki@mis.mpg.de, felix.gaisbauer@weizenbaum-institut.de, olbrich@mis.mpg.de\\[2mm]
}

\usepackage{amsmath}
\usepackage{amssymb}
\usepackage{tabularx}
\usepackage{booktabs}
\usepackage{bibentry}

\begin{document}

\maketitle

\begin{abstract}
We investigate the polarization of the German Twittersphere by extracting the main issues discussed and the signaled opinions of users towards those issues based on (re)tweets concerning trending topics. The dataset covers daily trending topics from March 2021 to July 2023. At the opinion level, we show that the online public sphere is largely divided into two camps, one consisting mainly of left-leaning, and another of right-leaning accounts. Further we observe that political issues are strongly aligned, contrary to what one may expect from surveys. This alignment is driven by two cores of strongly active users: influencers, who generate ideologically charged content, and multipliers, who facilitate the spread of this content. The latter are specific to social media and play a crucial role as intermediaries on the platform by curating and amplifying very specific types of content that match their ideological position, resulting in the overall observation of a strongly polarized public sphere. These results contribute to a better understanding of the mechanisms that shape online public opinion, and have implications for the regulation of platforms.
\end{abstract}

\section{Introduction}
\begin{figure*}[ht]
  \centering
  \includegraphics[width=\textwidth]{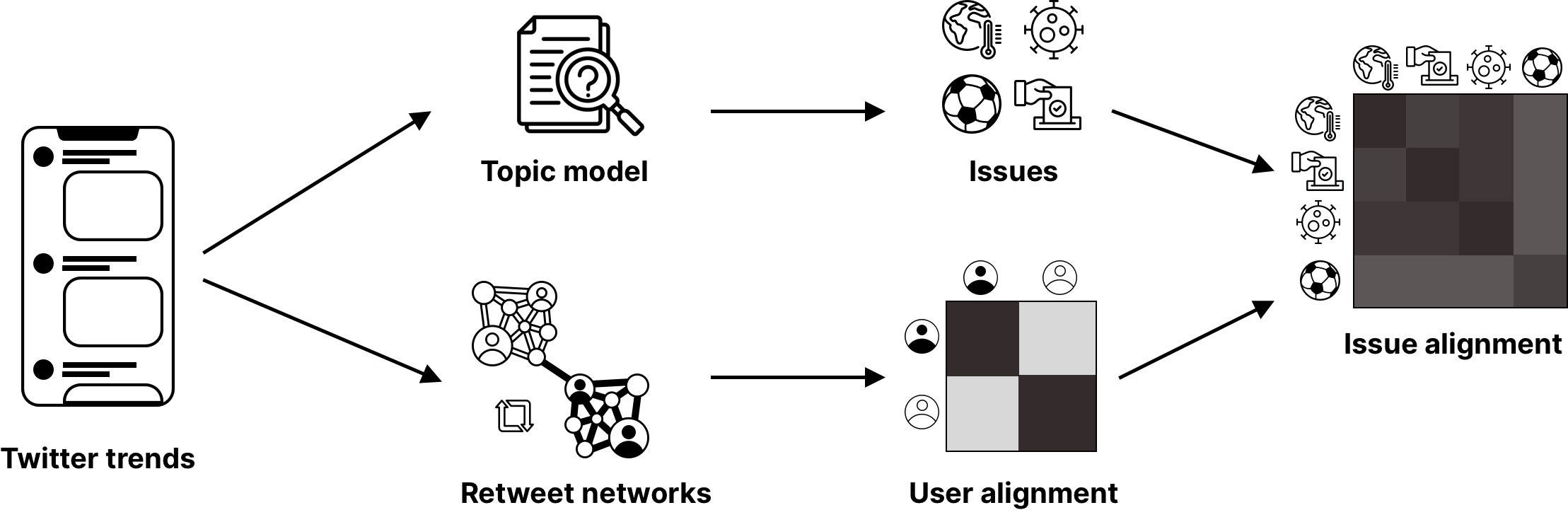}  
  \caption{Analysis pipeline. The raw text data of the tweets is processed in a topic model to extract the main issues discussed, which is then used to assign an overall issue to each trend. In parallel, retweet networks are constructed and clustered to extract the opinions of users for each trend. From this set of partitions, we extract the main actors (influencers and multipliers) and compute a \textit{user alignment} matrix that quantifies how systematically pairs of users appear in the same cluster across the dataset. Importing the issue labels from the topic model analysis allows us to compute the user alignment separately for each issue, which we then use to generate the issue alignment matrix that measures how similarly two issues sort these users into opinion groups.}
  \label{fig:pipeline}  
\end{figure*}
Social media platforms have become a significant part of the public sphere. Their rise in the past decades has reduced the gatekeeping role of traditional news outlets, as any private entity can now reach a wide audience on platforms like Twitter/X. While this democratization of the public sphere has been considered promising for political participation, it is nowadays often brought in connection with democracy-endangering phenomena like mistrust in institutions, rise of hate speech, misinformation and polarization.
While there is an increasing body of scientific evidence describing these phenomena \cite{Lorenz-Spreen2023}, a thorough understanding of their underlying mechanisms is still lacking. Empirical work is needed that thoroughly describes how users express and share opinions on social media platforms, and what role different actors may play in shaping the perception of (online) public opinion.\\
Existing empirical investigations on polarization on social media usually focus on individual issues, such as climate change \citep{Falkenberg2022}, Covid \citep{Quintana2022} or events, such as elections \citep{Conover2011a,Gaumont2018,Gaisbauer2021,Darius2022}. Only very few social-media based studies systematically analyze polarization at the level of multiple topics \citep{Chen2021,Salloum2024}. 
In such settings, opinions on a given set of issues can be correlated, which results in a phenomenon commonly described as \textit{issue alignment}. Apart from the two studies mentioned, issue alignment has so far mainly been investigated using survey data. For the USA, \citet{Baldassarri2008} show that, between 1972 and 2004, there is no strong evidence for issue alignment, but they find strong signals of issue partisanship, i.e. the correlation between the opinion towards an issue and party identification. Following a similar approach for the years after 2004, \citet{Kozlowski2021} show a growth in issue alignment, especially within topics related to civil rights and economy. In Europe, where multi-party-systems are likely to reduce the effects of issue partisanship, surveys show slightly higher polarization for certain issues like migration, but very little evidence of issue alignment \citep{Lux2022,Pless2023}.\\
What might follow from these empirical insights is that both polarization and issue alignment are restricted to smaller, but possibly opinion-leading groups. The notion of ``elite polarization'' has already been put forth in the 1960s: \citet{Converse1964} argues that only the \textit{political elite} is capable of strategically aligning issues and is therefore more ideologically polarized than the general electorate.
Yet, we argue that social media may have softened the distinction between elite and mass, at least on the digital platforms themselves. While traditional elites may still play an important role, laypeople now have the possibility to become highly influential \citep{Goodwin2023} by winning over the attention of a large number of users on platforms like Twitter/X or Facebook. As was pointed out in \citet{Papacharissi2017}, digital platforms \textit{promise} this potential, but do not \textit{guarantee} it. What we undoubtedly observe on social media, however, can be thought of as a new distinction driven by stark differences in user activity and engagement. While the majority of users are mainly content consumers \citep{Hughes2019}, the most active users play key roles in shaping the online public sphere. Depending on the affordances of the studied platform, this activity can be further subdivided into different types. On Twitter/X, retweets are the central type of engagement through which information is relayed. That is, not only content \textit{creation}, but additionally, content \textit{distribution} is of central importance on the platform.
We hence assume that two types of ``power users'' play a significant role in shaping public debate on social media. On the one hand, most of the content is provided by \textit{influencers} -- prominent users reaching a large audience on the platform. Previous work on social media influencers has investigated their impact on information diffusion \citep{Watts2007,Bakshy2011}, their role in potentially furthering political education and participation \citep{Riedl2021,Dekoninck2022,Sehl2023}, spreading of false news \citep{Starbird2023} or hate speech \citep{Stewart2023}. Here, we consider influencers as the content creators with the ability to reach a wide audience with their messages. Yet, in social media environments, information diffusion is no longer really a one-to-many process. Communication is, on the contrary, often intermediated. Information flows are therefore no longer fully characterized by a dyadic relationship between two actors (the source, which distributes its own content, and a consumer), because a third element -- an intermediary -- can be positioned between them \citep{friemel2023public}. The intermediary hereby produces a service to the source, the recipient, or both. In its most basic form, this can be thought of as curation: the reception, selection, arrangement, aggregation and redistribution of content produced by other actors \citep[p. 96]{friemel2023public}. Empirically, previous work investigated the role of intermediaries in political information flow \citep{Hemsley2019,Yoon2022,Liang2023}, though, to the best of our knowledge, not in relation with polarization and issue alignment. The present work aims to fill this gap. In the following, we refer to users acting most actively as intermediaries as \textit{multipliers}.
We apprehend these concepts quantitatively by interpreting users that are retweeted the most as influencers, and users that retweet the most as multipliers.\footnote{These concepts are not mutually exclusive: An influencer could also retweet massively and act as a multiplier. We will show empirically that there is no significant overlap between the groups (see Fig.~\ref{fig:appendix:infmul_overlap}), and that multipliers do indeed enjoy a broader audience that the average user, i.e. they are not `broadcasting into the void' (see Fig.~\ref{fig:ccdf}).}\\
In this work, we shed light on the role these two user types play in the emergence of polarization and issue alignment on social media. Analyzing a large dataset of tweets collected from trending topics in Germany between March 2021 and July 2023, we address the following research questions: What are the topics that trend in the German Twittersphere? Which ones spark (more or less) polarized discussions? Do we find evidence for issue alignment across these topics? What is the potential role of different types of power users? Approaching these questions using a combination of topic modeling and network analysis, we show that there is strong evidence for issue alignment across political topics, in the sense that those issues sort users into similar opinion groups. These groups are characterized by a division into left and right, which reflects the findings of previous work on the German Twittersphere \citep{Gaisbauer2021,Darius2022}. We hypothesize that this division is driven by two types of hyperactive users, influencers and multipliers. The former generate the majority of circulated content, while the latter select and aggregate these contents into ideologically consistent bundles. This combination of user behaviors leads to the observation of a polarized and aligned online public sphere.
\section{Data}
The dataset includes tweets on trending topics over the course of more than two years. More specifically, we collected tweets from 2021-03-29 to 2023-07-12 according to the following scheme: at the beginning of each day, we launched a script that collects the current ``trending topics'' (from now on referred to as ``trends'') in Germany using the Twitter Trend API (v1). By default, trends are personalized based on the account's Twitter/X usage. One can, however, disable the personalization by setting a specific location from which to draw the trending topics, which then yields ``popular topics among people in a specific geographic location'' \citep{X/Twitter2025}. We re-ran the script every 15 minutes. At the end of each day, we counted the number of times each trending topic appeared during the day and kept the top 5 most frequent ones. This gave us a proxy of the five most important trending topics for that day. We then used the Twitter Search API (v1) to collect German-speaking tweets using the exact trend keyword as a query on the day it trended and the day after (48hrs). All the tweets were collected using a single Twitter API key, collecting tweets for maximally 24 hours every day. Each collected trend is considered one sub-dataset. This results in 3007 trending topics collected. We merged trends if the exact same phrase occurs within a time window of 1 day. After merging, there are 2693 sub-datasets. These contain a total of 19,105,532 tweets.
\section{Methodological Challenges}
\subsection{Extracting Issues from Tweets}
\label{subsec:method_topicmodel}
In order to extract the main issues discussed on German Twitter trends, we perform topic modeling on the collected tweets. We restrict the corpus to original tweets that have either been liked or retweeted at least once. This results in a corpus of approx. 4 million tweets. Since tweets are short, idiosyncratic texts, traditional topic modeling techniques such as LDA have a tendency to perform poorly \citep{Hong2010,Egger2022}. We therefore use a sentence-embedding based approach packaged in the Python library BERTopic \citep{Grootendorst2022}. The method consists of embedding the documents into a space in which distances reflect semantic similarity. Clusters in that space are then characterized by topical coherence. The task of topic extraction is therefore cast to a task of clustering in a semantic space. The space is usually given by pre-trained sentence-embedding models. Since these spaces typically span several hundred dimensions, the curse of dimensionality sets in and clustering needs to be performed on a lower-dimensional representation. These lower-dimensional representations are obtained using UMAP \citep{McInnes2020}. In a final step, a density-based clustering algorithm, typically HDBScan \citep{McInnes2017}, is used to uncover the documents that belong to the same topic. Then, the topics are hand-labeled based on the most relevant words for each respective cluster computed using a tf-idf score. The density-based partition is hierarchical, which allows us to merge sub-topics into larger themes or issues. Please refer to Sec.~A.1 for a detailed description of the parameters and sentence-embedding model used.\\
We apply this scheme on a representative sub-set of the corpus, keeping the top 50 most retweeted tweets of each trend. This results in about 100,000 documents, for which the above mentioned computations are not costly. We compute the topic model on this sub-corpus and generate a set of document-topic pairs, which we then treat as hand-annotated data consisting of a chunk of text, the corresponding sentence-embedding vector and the label given by the topic model. Using the vector-label pair, we train a logistic classifier that learns the features in the semantic space that predict a given topic. The weights of this classifier are then used to infer the documents of the full corpus of 4 million tweets. This second part of the method is known as "supervised topic modeling" \citep{Grootendorst2024}. Finally, once we have assigned a topic label to each tweet, we use a majority rule to assign a topic to each trend based on the topic distribution of its tweets. 
\subsection{Extracting Opinions from Tweets}
\label{subsec:rtn}
\begin{figure}[t]
  \centering
  \includegraphics[width=.48\textwidth]{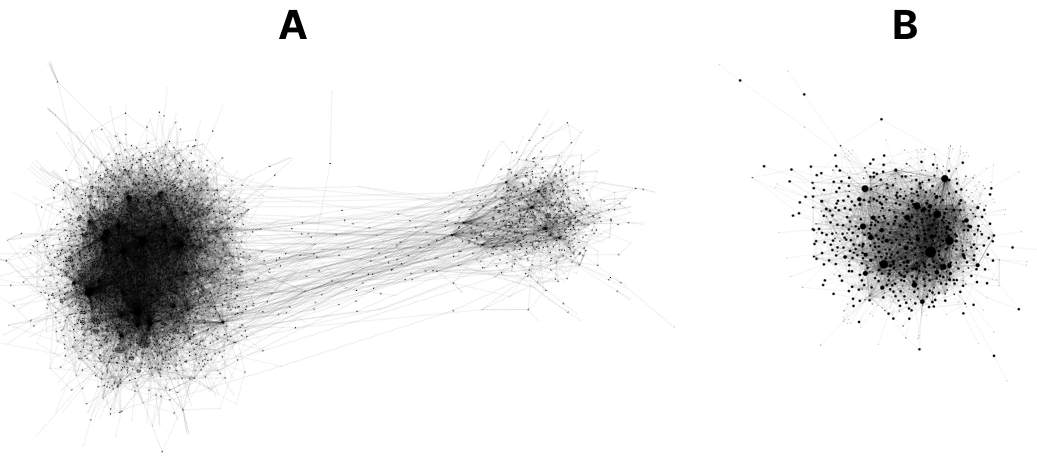}  
  \caption{Force-directed layout representation of two retweet networks. \textbf{A} shows a retweet network from a polarized trend, \textbf{B} one from an unpolarized trend.}
  \label{fig:rtn-example}  
\end{figure}
Each trend in our dataset represents a discussion over an issue identified using the topic model. Our aim is to extract the different opinions present in each discussion. Drawing on previous work \citep{Conover2011a,Gaumont2018,Gaisbauer2021}, we use the retweet network as the central object of analysis. Each node is a user, and a directed link is drawn from $i$ to $j$ if $i$ retweets $j$. Since retweets are assumed to be endorsements \citep{Boyd2010}, we can interpret densely connected groups of nodes in the retweet network as opinion groups in the underlying debate. Previous work suggests that the high-level structure of political retweet networks is bipolar \citep{Conover2011a,Gaisbauer2021}. Spatializing the retweet networks in the dataset using a force-directed layout algorithm \citep{Jacomy2014}, which tends to place groups of densely connected nodes closer to each other, shows that the majority of trends in our dataset generate retweet networks with either two or one cluster. Figure~\ref{fig:rtn-example} shows an example for each case. Our starting assumption is therefore that a trend can either be polarized or not, or in other words, there can be either two opposing opinion camps or one single consensual camp. Conceptually, we treat the retweet network as an opinion poll. For each polarized issue, a user can either take the stance $-1$ or $1$, or they can abstain from responding. The retweet network from tweets about the Covid vaccine, for instance, divides into two camps: proponents and opponents of the vaccine. In order to operationalize these assumptions, we systematically assess whether a given retweet network is polarized or not, i.e. whether it divides in two camps or not. For this, we infer a degree-corrected stochastic block model (SBM), which allows us to infer whether the most likely generative model of the observed graph follows a block structure. Since we expect from manual inspection of the force-layouts that either one or two blocks as the most likely generative model, we infer the SBM under the constraint of $N_{blocks} \in \{1,2\}$ using a minimum description length procedure implemented in the Python library graph-tool \citep{peixoto_graph-tool_2014}. If the 1-block model has a smaller description length, then we assume the trend not to be polarized. The opinion groups we seek, characterized by densely connected nodes that are more strongly connected within their cluster than across the network, correspond to \textit{assortative communities}. The SBM is also able to uncover disassortative communities, such as groups of structurally equivalent nodes or core-periphery structures \citep{Liu2023}. These communities typically do not correspond to opinion groups, but may for instance group together hubs from each opinion camp. In order to make sure that the SBM produced assortative communities\footnote{In principle, one could force the SBM to recover assortative communities using the planted partition model \citep{Zhang2022}. In practice, however, this model was often not able to faithfully recover the clusterings observed in the force-directed layout.}, we use the previously computed force-layout spatialization of the graph in which the assortative clustering becomes visually apparent \citep{Noack2009}. We compute the quality of the SBM clustering using a silhouette score on the 2D embedding given by the force layout. If the score is higher than 0.4, we keep the SBM-based cluster assignment, otherwise assume that there is only one cluster. In a final step, the clustering-embedding pairs are hand-checked for each network and re-computed if the embedding differs strongly from the clustering. This procedure generates a set of partitions, one for each graph, that we use to measure the alignment of users and issues. Please refer to Sec.~A.2 for a detailed description of the clustering procedure.
\subsection{Measuring User Alignment}
\begin{table*}[ht]
\begin{tabular}{lrrrrr}
\toprule
Topic & $N_{\mathrm{tweets}}$ & Retweet share & $N_{\mathrm{trends}}$ & $N_{\mathrm{trends}}$ with $|V| \geq 50$ & Polarized trends share \\
\midrule
German Politics & 2276715 & 0.82 & 338 & 305 & 0.71 \\
Covid & 1862823 & 0.78 & 250 & 217 & 0.70 \\
Ukraine & 1701862 & 0.79 & 205 & 177 & 0.66 \\
Climate Change & 1461830 & 0.80 & 157 & 133 & 0.60 \\
Sports & 920527 & 0.65 & 514 & 180 & 0.36 \\
Greetings and Holidays & 822367 & 0.48 & 225 & 94 & 0.39 \\
Foreign Politics & 768439 & 0.79 & 97 & 73 & 0.60 \\
Democracy & 709776 & 0.77 & 62 & 59 & 0.69 \\
Journalism/Media & 497964 & 0.82 & 50 & 44 & 0.64 \\
Police & 467491 & 0.85 & 61 & 55 & 0.65 \\
Social Politics & 463404 & 0.77 & 73 & 58 & 0.52 \\
Social Media & 416302 & 0.75 & 71 & 34 & 0.41 \\
Gender/LGBTQ & 408078 & 0.77 & 71 & 50 & 0.70 \\
Right-Wing Extremism & 347301 & 0.82 & 17 & 16 & 0.81 \\
Energy & 318021 & 0.77 & 43 & 36 & 0.72 \\
Pop Culture & 269714 & 0.59 & 175 & 55 & 0.31 \\
Racism & 244734 & 0.78 & 41 & 36 & 0.64 \\
Migration & 206589 & 0.83 & 17 & 15 & 0.60 \\
Religion & 185335 & 0.72 & 32 & 20 & 0.55 \\
Mobility & 138568 & 0.73 & 19 & 15 & 0.33 \\
Antisemitism & 137434 & 0.83 & 14 & 12 & 0.67 \\
Drug Legalisation & 132207 & 0.74 & 21 & 16 & 0.25 \\
Gaming & 113756 & 0.65 & 64 & 9 & 0.11 \\
Music & 49365 & 0.55 & 60 & 14 & 0.21 \\
Abortion & 16567 & 0.76 & 4 & 3 & 0.67 \\
\bottomrule
\end{tabular}
\caption{Main issues discussed in German Twitter trends from March 2021 and July 2023.}
\label{tab:topicmodel}
\end{table*}
\begin{figure*}[ht]
  \centering
  \includegraphics[width=\textwidth]{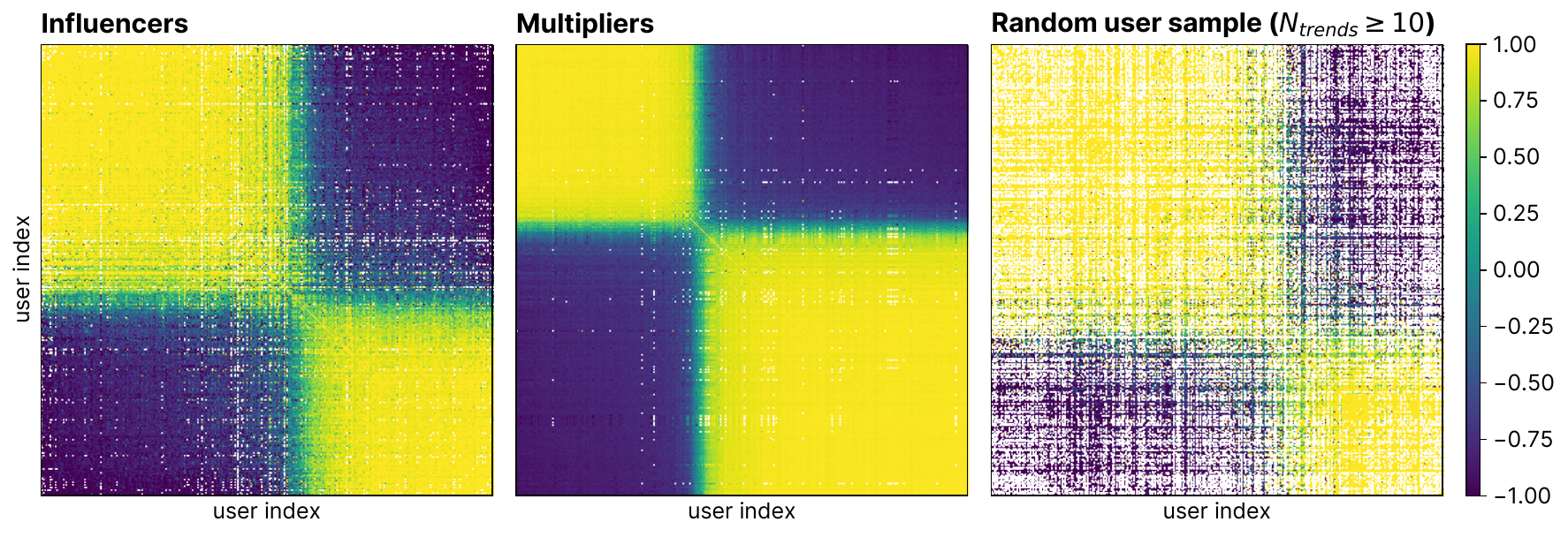}
  \caption{User alignment for influencers, multipliers and a random sample of users. We compute the pairwise user alignment for the top 1000 influencers (highest in-degree), top 1000 multipliers (highest out-degree) and a set of 1000 randomly sampled users. The matrices are sorted based on a hierarchical linkage clustering. If users do not participate in any same trend, the matrix field is left white. For each user set, there is a left and right-leaning cluster. In between, there are users that retweet and are retweeted by both opinion groups. Note that multipliers are more strongly aligned than the other groups and the right-leaning cluster is significantly larger. The matrix for the random sample shows that the division into the two camps exists here too, but the common participation in retweet networks is lower.}
  \label{fig:useralignment}
\end{figure*}
\begin{figure*}[ht]
  \includegraphics[width=\textwidth]{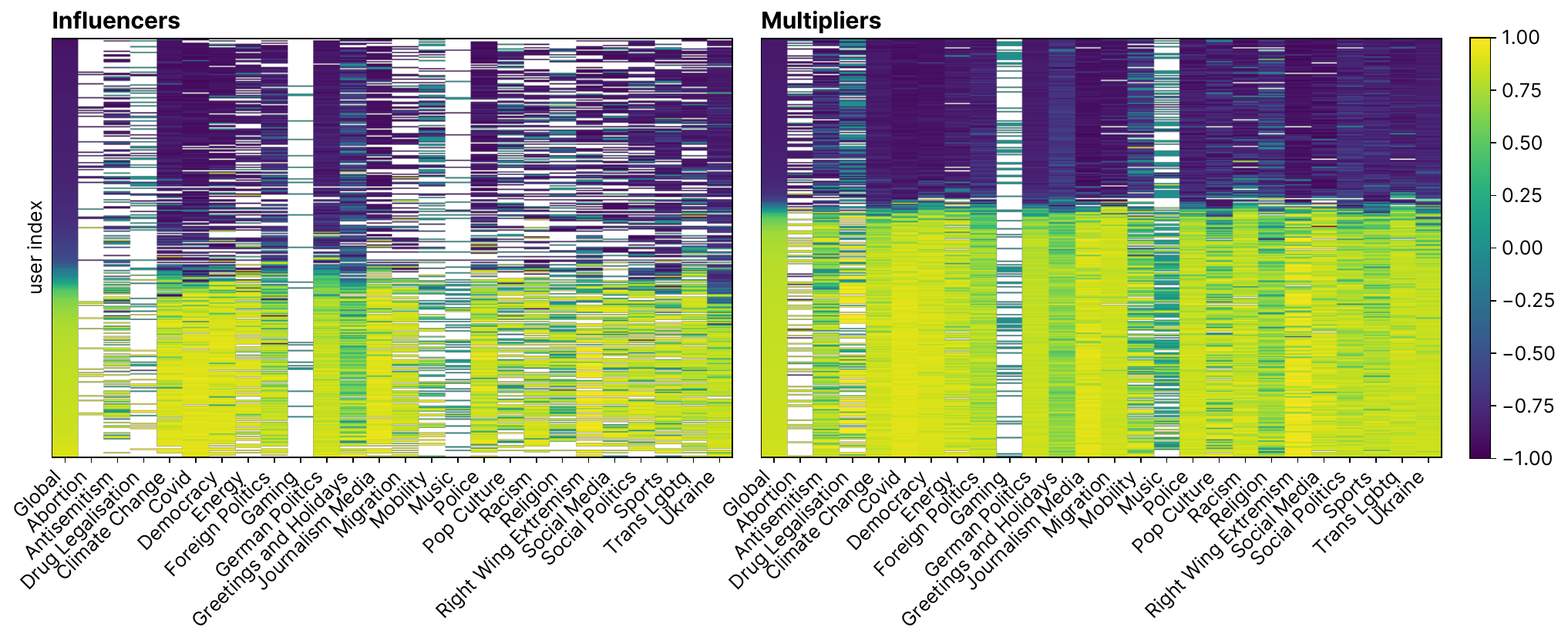}
  \caption{Global and topic-wise cluster membership score for influencers (left) and multipliers (right). A membership score of $-1$ means that the user belongs to the left-leaning cluster (blue color), $+1$ to the right-leaning cluster (green color). Blank lines in the matrix mean that the user did not participate in any retweet network associated to the given topic. We observe that multipliers are more active across topics than influencers.}
  \label{fig:user-topicalignment}
\end{figure*}
\begin{figure*}[ht]
  \includegraphics[width=\textwidth]{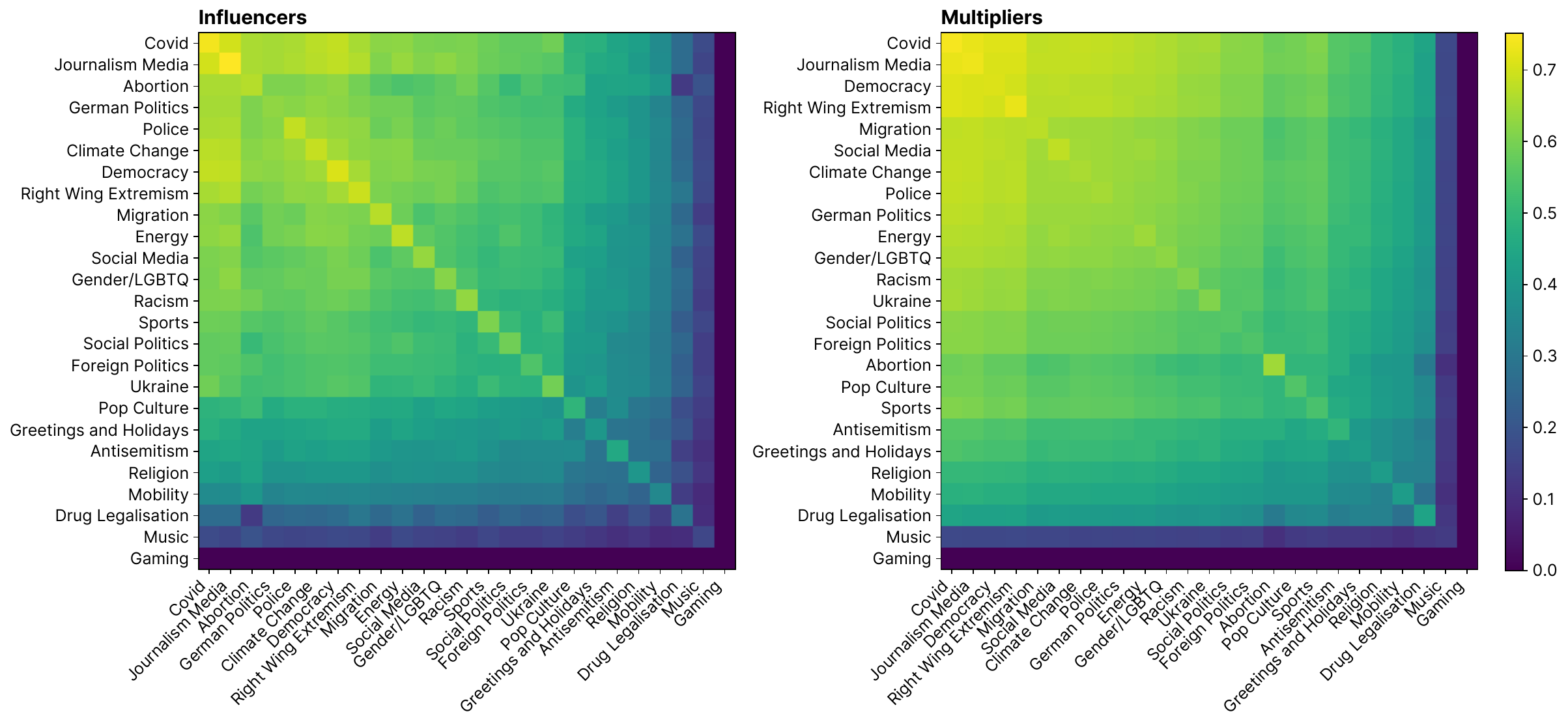}
  \caption{Issue alignment for influencers (left) and multipliers (right). Both matrices are sorted according to optimal leaf ordering. For both user groups, we observe a strong issue alignment across topics, except for Music and Gaming. Multipliers exhibit a stronger issue alignment than influencers. There is no apparent cluster structure in either matrix.}
  \label{fig:issuealignment}
\end{figure*}
\label{subsec:useralignment}
There are $N_{\mathrm{trends}}$ trends in our dataset. For each trend $k$, we encode the cluster assignment of users in the vector $c_k$ as follows:
\begin{equation}
    c_k(i) = \begin{cases} +1 & \text{if user } i \text{ is in cluster $1$}\\ -1 & \text{if user } i \text{ is in cluster $-1$}\\  0 & \text{if user } i \text{ is not in the trend}\\\end{cases}
\end{equation}
We define the trends index set $\mathcal{I} \in \{1,...,N_{\mathrm{trends}}\}$. Each trend belongs to one topic. This allows us to define the topic index set $\mathcal{I}_T$ that contains all the trend indices that are part of topic $T$. 
We then compute the user alignment $\alpha(i,j) \in [-1,1]$ between two users $i$ and $j$ on the subset of $m(i,j)$ trends in which they both participate. 
\begin{equation}
  \label{eq:useralignment}
  \begin{aligned}
    \alpha(i,j)=\frac{1}{m(i,j)} \sum_{k \in \mathcal{I}} c_k(i)c_k(j)
  \end{aligned}
\end{equation}
where $m(i,j)$ is defined as
\begin{equation}
  \label{}
  \begin{aligned}
    m(i,j)=\sum_{k \in {\mathcal{I}}}|c_k(i)c_k(j)|
  \end{aligned}
\end{equation}
If two users are always in the same cluster, then $\alpha(i,j)=1$, if they always oppose each other, then $\alpha(i,j)=-1$. If a pair of users were randomly assigned to the cluster their average alignment would be $0$. 
Similarly, we define the topic-wise user alignment $\alpha_T$
\begin{equation}
  \label{}
  \begin{aligned}
    \alpha_T(i,j)=\frac{1}{m_T(i,j)} \sum_{k \in \mathcal{I}_T} c_k(i)c_k(j)
  \end{aligned}
\end{equation}
where $m_T(i,j)$ is defined as
\begin{equation}
  \label{}
  \begin{aligned}
    m_T(i,j)=\sum_{k \in {\mathcal{I}_T}}|c_k(i)c_k(j)|
  \end{aligned}
\end{equation}
There are 700,025 users in the dataset. However, most of them only participate in a small number of trends. Our data follows the usual pattern that in social networks, namely that different users play more or less important roles in the networks. Degree distributions typically follow something that resembles a heavy tailed distribution (e.g. a power law): a large number of users have a low degree (both in and out), while a few users (so-called hubs) have a very high degree. Since the networks we observe are directed, we can use both the in-degree (the number of times a user got retweeted) and the out-degree (the number of times a user retweeted other users' contents) to distinguish between two user types: \textit{influencers}, those users with the highest in-degree, and \textit{multipliers}, those with the highest out-degree. 
\subsection{Clustering Aligned Users}
For both influencers and multipliers combined, we first generate the user alignment matrix by computing all pairwise alignments $\alpha(i,j)~\forall~i,j$. We then cluster the resulting matrix using hierarchical linkage clustering, which yields two clusters. Close examination of the most influential accounts and their (re)tweet activity allowed us to label the clusters as a left-leaning and right-leaning cluster, which we denote by $l$ and $r$. 
This division is encoded in the global cluster vector $c(i) \in \{l,r\}$.
We can now use these ``global'' camps made of influencers and multipliers as anchors and compute, for any given user, how well they align with these camps, i.e. how often they appear in the same retweet network cluster as any user in the global camps.
If user $i$, for instance, more often find themselves in the same cluster as users from $l$, then their Twitter/X activity is more left-leaning. We quantify the degree of membership to the global camp $k$ using the score $\nu(i,k) \in [0,1]$ defined as
\begin{equation}
    \nu(i,k)= \frac{\sum_{j \in J_i} \alpha(i,j)~\delta(c(j),k)}{\sum_{j \in J_i} \delta(c(j),k)}
\end{equation}
where $J_i$ denotes the subset of influencers and multipliers which occur jointly with $i$ at least in one trend and ${\delta(i,j)=1}$ if $i=j$ and $0$ otherwise. The overall cluster membership score $\mu(i) \in [-1,+1]$  is defined as 
\begin{equation}
    \mu(i) = \frac{\nu(i,r) - \nu(i,l)}{2}
\end{equation}
This score is $-1$ if the user is maximally aligned to the left-leaning cluster, $+1$ is they are maximally aligned to the right-leaning cluster and 0 if they are in between.
\subsection{Measuring Issue Alignment}
\label{subsec:issuealignment}
Previous approaches to measuring issue alignment on social media relied on information-theoretic measures of partition similarity \citep{Chen2021,Salloum2024}. Measures such as Normalized Mutual Information can be computed on the set of overlapping users between two clusterings. Sometimes, this overlap can make up only a small fraction of the nodes of the two graphs, which makes the resulting alignment scores difficult to interpret. Furthermore, these scores typically need to be adjusted for random cluster assignment, and defining a sensible null-model for the assignment of users to retweet clusters is non-trivial. The same holds for count-based partition similarity measures like the Rand score. We therefore propose a different approach, based on the user alignment defined in the previous section, and refer the reader to the Appendix~\ref{appendix:alignment_measures} for an in-depth comparison of different alignment measures. In order to compare the way different topics align users in different (or similar) ways, we define the degree of membership of user $i$ to global camp $k$ for each topic $T$ separately:
\begin{equation}
    \nu_T(i,k)= \frac{\sum_{j \in J_i} \alpha_T(i,j)~\delta(c(j),k)}{2\sum_{j \in J_i} \delta(c(j),k)}
\end{equation}
The topic-wise cluster membership score is then defined as
\begin{equation}
    \mu_T(i) = \frac{\nu_T(i,r) - \nu_T(i,l)}{2}
\end{equation}
The alignment $\tau(T_1,T_2) \in [0,1]$ between two topics $T_1$ and $T_2$ is then defined as 
\begin{equation}
    \tau(T_1,T_2) = \frac{1}{n}\sum_i^{n} \mu_{T_1}(i)\mu_{T_2}(i)
\end{equation}
where $n$ is the number of users for which the alignment is computed. Intuitively, we measure how strongly the cluster assignments extracted from retweet networks of one topic match the assignments of another topic. If two topics $T_1$ and $T_2$ sort users into the exact same opinion groups, then $\tau(T_1,T_2) = 1$.
\section{Results}
\subsection{Topics}
\label{subsec:topicmodel}
Using the BERTopic-based topic modeling procedure described in Sec.~\ref{subsec:method_topicmodel}, we characterize the discussion space in the German Twittersphere by extracting the main issues that generate trending discussions. Combined with the retweet-network based opinion extraction described in Sec.~\ref{subsec:rtn}, we estimate which issues sparked more or less polarized discussions by computing the share of retweet networks best described by two assortative blocks. The results are summarized in Tab.~\ref{tab:topicmodel}. We see that the topics generating the most tweets are related to political issues. Among the most polarized topics, we find \textit{Right-Wing Extremism}, \textit{Energy}, with discussions around nuclear energy, electric vehicles or heat pumps, \textit{German Politics}, with discussions around elections (e.g ``\#ltwnrw22''), specific parties (e.g. ``CDU und FDP'') or politicians (e.g ``\#Laschet''), \textit{Covid}, with issues like lockdowns (e.g. ``\#harterLockdownJetzt''), vaccination (e.g. ``\#Impfneid'') or masks (e.g. ``\#DieMaskeMussWeg'') and \textit{Gender/LGBTQ} which mainly contains discussions around women's (e.g. ``\#InternationalWomensDay'') and trans rights (e.g. ``\#TransDayOfVisibility''). The least polarized issues are \textit{Gaming}, which mainly consists of discussions around new video games, and \textit{Music}. The topic of \textit{Drug Legalisation}, mainly related to cannabis, is also fairly unpolarized, which is due to the fact that the opponents of cannabis legalisation are not very vocal in the collected trends. \textit{Sports}, which in the German Twittersphere is mainly soccer-related, is another example of a rather unpolarized topic, since the supporters of both teams usually retweet the same hubs. The topic that generated the highest number of original tweets is the one related to \textit{Greetings and Holidays}. It contains tweets like ``Happy Sunday, everyone!'' or ``Merry Christmas!''. Its share of polarized trends is still close to 40\%, which is due to the fact that the users tend to retweet each other's wishes within their ideological cluster. The \textit{Journalism/Media} topic is related to the role of traditional media and new alternative media, integrity of journalists and satirical news formats. Looking at the fifth column in Tab.~\ref{tab:topicmodel}, we see that some topics, even though they may generate many trends, rarely generate enough retweets to spark controversy: \textit{Pop Culture}, \textit{Music}, \textit{Gaming}, \textit{Social Media} are examples of topics in which less than half of the retweet networks are made up of less than 50 users. 
\subsection{User Alignment}
\label{subsec:results_useralignment}
Each topic has a corresponding set of retweet networks that cluster the user base into more or less similar clusters. Using the measure of user alignment between two users $\alpha(i,j)$ defined in \eqref{eq:useralignment}, we quantify how systematically two users sort themselves into the same cluster across the dataset. If two users $i$ and $j$ are always aligned with respect to their opinion cluster, then $\alpha(i,j)=1$.\\ 
Figure~\ref{fig:useralignment} shows the alignment matrix for influencers, multipliers and a random sample of users that participate in retweet networks of at least 10 different trends. In each case, we re-ordered the matrix using the cluster membership score of each user $\mu(i)$. We consistently observe a division into two clusters. Manually examining the accounts and their (re)tweet activity shows that, in all three cases, we find a left-leaning and right-leaning cluster. In between these two clusters, there are users that retweet and are retweeted by both opinion groups. The influencers in the ideological camps mainly consist of politicians, journalists, activists, bloggers and (alternative) media outlets. The influencers bordering the two camps mainly consist of newspaper outlets, center right politicians and football-related accounts. The multipliers mostly correspond to accounts that are not public figures, they are less notorious and usually not identifiable from examining their Twitter profile. Some of them mention their ideological leaning in their description. Note the size differences between the left-leaning and the right-leaning clusters. For multipliers, the right-leaning cluster has more users, while for the two other cases, the left-leaning cluster is larger. The alignment of the random user sample exhibits the same left-right structure as the two high-activity users.\\
To further characterize the activity of influencers and multipliers, we compare their distributions of number of trends in which they participate $N_{\mathrm{trends}}$ and number of followers $N_{\mathrm{followers}}$ to those of all users in the dataset. The result is shown in Fig.~\ref{fig:ccdf}. Multipliers are more active in terms of numbers of trends than all other users, including influencers. Influencers have more followers than other users, but multipliers still have more followers than average users.
\subsection{Issue Alignment}
\label{subsec:results_issuealignment}
Figure~\ref{fig:user-topicalignment} shows the global and topic-wise cluster membership score for the top 1000 influencers and multipliers. Multipliers are more active and more consistently aligned across topics. The topics \textit{Abortion}, \textit{Drug Legalisation}, \textit{Gaming} and \textit{Music} seem to attract different users than the other topics, which is why the membership scores are very sparse. Comparing influencers and multipliers at the level of individual topics, we see some more nuanced differences: 
\textit{Migration} is more strongly aligned with other issues for multipliers than for influencers. Figure~\ref{fig:user-topicalignment} shows that more right-leaning than left-leaning influencers take part in migration-related discussions. Some of those left-leaning accounts that took part find themselves in the opposing camp. In order to explain this, we closely examined (re)tweets by those influencers that ``fell out'' of their cluster using the Python library twitter-explorer \citep{Pournaki2021} to visualize the graphs interactively and read the underlying tweets. The reduced alignment can be explained by different factors. First and foremost, the topic of migration is traditionally more salient on the right. This leads to the fact that left-leaning newspapers reporting on events related to migration or migrants tend to be retweeted more by right-leaning accounts. Similarly, some left-leaning influencers that expressed themselves in some migration-related trends were retweeted by the right, ``pulling'' them towards the opposing cluster. Interestingly, this phenomenon does not happen for multipliers, where the \textit{Migration} topic is among the most aligned topics. 
Looking more closely at trends related to \textit{Ukraine}, we see that the topic had a tendency to re-arrange the left and right camps. This can be explained by the fact that certain right-wing influencers did not share the pro-Russian stance of the rest of their camp. This effect is, again, less visible when looking at multipliers. Both examples show that multipliers align issues more consistently than influencers.\\
In order to measure the pairwise alignment between topics, we computed the topic alignment matrix shown in Fig.~\ref{fig:issuealignment}. The diagonal entries of the matrix show how strongly the issue is aligned internally and measure how similar the opinion clusters of retweet networks are within that same topic. The off-diagonal entries show how aligned two topics are. The matrices are sorted according to optimal leaf ordering, which sorts the issues by most aligned (top left) to least aligned (bottom right). We observe a very high issue alignment across topics, except for \textit{Music} and \textit{Gaming}, which are weakly polarized topics that do not include a large number of influencers and multipliers. We cannot observe any significant cluster structure that would hint towards groups of strongly aligned issues, misaligned across groups, in either of the two issue alignment matrices. We rather see a gradual difference, going from strongly aligned issues like \textit{Covid, Journalism, Abortion} or \textit{German Politics} towards less aligned topics like \textit{Social Politics}. Finally, comparing the issue alignment of influencers and multipliers, we see that the latter align issues more strongly than the former.
\section{Discussion}
\begin{figure}[t]
    \centering
    \includegraphics[width=.48\textwidth]{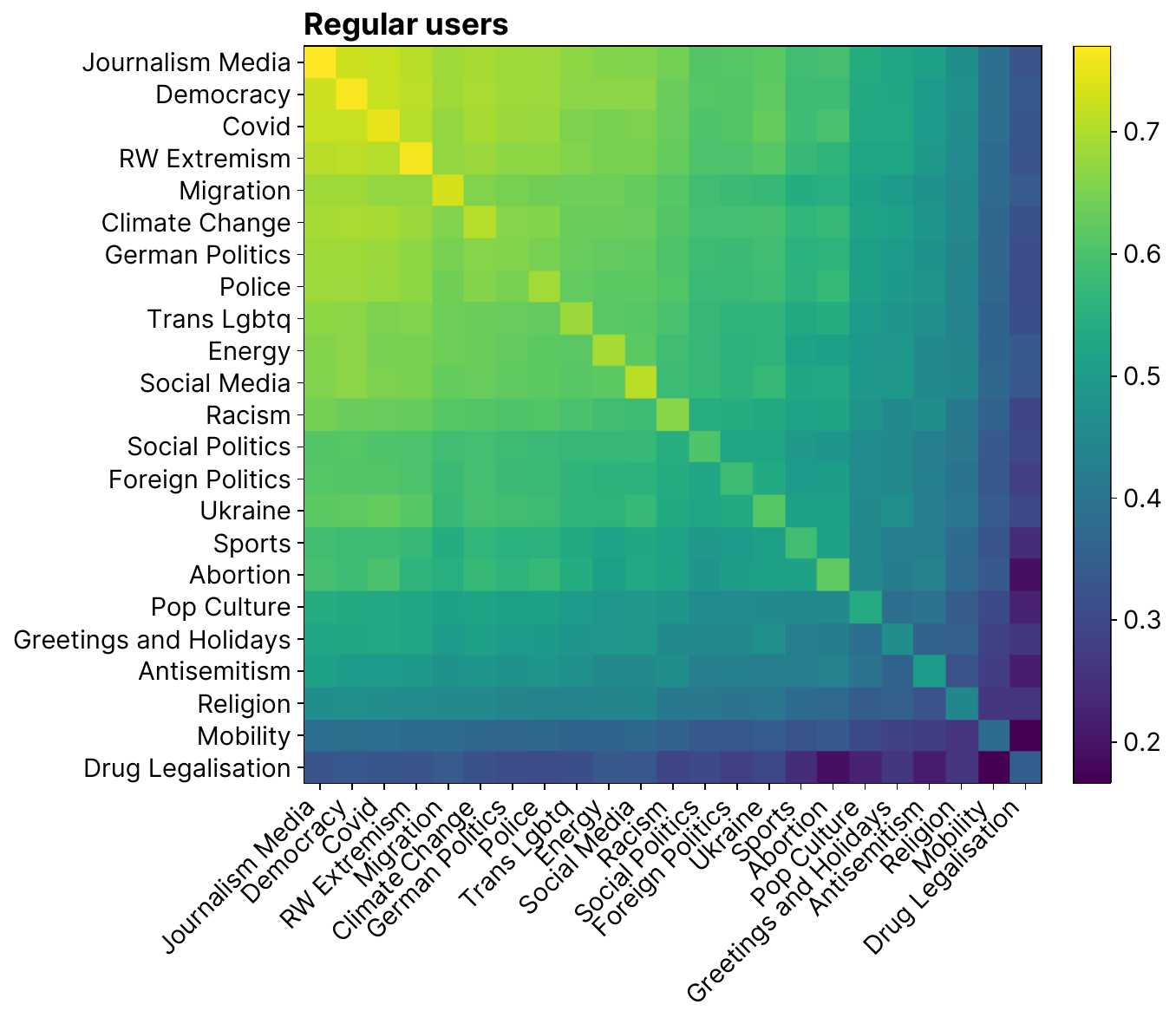}
    \caption{Issue alignment of regular users that participate in at least 10 topics ($N=13440$). We see strong alignment as for influencers and multipliers, and that traditionally left-right dividing topics like social politics are less aligned to the global division than topics related to the role of journalism and news, Covid or right-wing extremism.}
    \label{fig:issuealignment-regular}
\end{figure}    
In this work, we have analyzed a substantive sample of the German Twittersphere between March 2021 and July 2023. Extracting the main trending topics from this period, we have found further evidence that Twitter/X is a highly politicized platform where the majority of tweets in trending topics are related to political issues, as can be seen in Tab.~\ref{tab:topicmodel}. Our starting point for the analysis was the retweet network, which, following previous literature \citep{Boyd2010,Conover2011a,Gaumont2018,Gaisbauer2021,Darius2022}, we considered a network of endorsement. Clusters in such networks can be interpreted as opinion clusters if the underlying discussion is political. Following this approach, we found that the political topics largely divide the user base into two clusters. Figure~\ref{fig:useralignment} shows that this does not only hold for the two subsets of hyper-active users (influencers and multipliers), but also for a randomly sampled set of users that took part in more than 10 trends. We can therefore report a significant structural polarization in the German Twittersphere. Strikingly, this polarization is consistent across issues, and the topic-wise configuration of clusters demonstrates evidence of high issue alignment across issues. Looking more closely, we still observe some differences in alignment between issues, which allows us to speculate about the nature of the observed division. Looking at the alignment matrix in Fig.~\ref{fig:issuealignment}, we find a core of strongly aligned topics related to Covid, the role of journalism and alternative media, and German politics, whereas topics corresponding to traditional left-right fault lines like social politics show less alignment. This leads us to assume that the observed clusters correspond to a new cleavage previously hypothesized in the political science literature between ``national populists'' (sympathizers of the right-wing party AfD) and ``liberal universalists'' \cite{Kriesi2006,Bornschier2010}. Two examples of topics that deviate slightly from this global alignment scheme are related to migration and the Russian invasion of Ukraine. The issue of migration is traditionally owned by right-leaning parties in Germany. A qualitative analysis of the corresponding retweet networks shows that users from the left-leaning camp are retweeted by the right-leaning camp more often than usual, even when they tweet neutrally about this topic. This forces them into the opposing cluster. In the case of the conflict in Ukraine, some right-leaning influencers do not share the general pro-Russian stance of their ideological allies, which leads them out of their usual cluster. Aside from these deviations, we observe high alignment across political issues, which goes against previous findings from survey-based research \citep{Baldassarri2008,Kozlowski2021,Pless2023}. How can we explain this difference? In the case of social media presented here, the measurement is different, and, more importantly, the types of users that are ``surveyed'' are different. While traditional surveys aim to reach a representative sample of the population, Twitter/X is known to be younger and more politicized than average \cite{Hughes2019}. Furthermore, the context in which users are expressing their opinion is different on social media than in surveys. Taking the example of migration, a general survey question about the benefits of migration may result in different answers from the same respondent than an inflammatory post about an immigrant committing a terrible crime. \citet{Mau2023} call this phenomenon ``trigger points'', and we hypothesize that trending topics on social media may mainly consist of such triggering elements, since they are likely to generate stronger engagement than moderate descriptions of the same topic.\\ 
Connecting our findings to previous work, we report that the alignment we observe is similar to the one \citet{Chen2021} have found in the Finnish Twittersphere. They show that the topic of \textit{Climate Change} is aligned with \textit{Migration}, and that both are aligned to the new political cleavage discussed above. More recently, \citet{Salloum2024} have shown a consistently high alignment across five issues (climate, migration, social security, economic policy and education) for ``elites'' on social media. These findings, along with other previously cited work on retweet networks, hint towards the fact that the division into antagonistic camps across issues is a general trend on social media. The natural next question is, what drives this division? We claim in this work that it is driven by two types of users that we identify using the in- and out-degrees in the networks. Influencers are opinion leaders, they generate most of the retweeted content and use the platform to spread their political opinion. Even though social media platforms might facilitate the possibility of generating influencers, they are not necessarily specific to the platforms. Multipliers, on the other hand, are more specific to social media. For one, they act as content curators for the users that follow them. By selectively retweeting specific content, they bundle together political messages from influencers and create an ideologically consistent package. This expresses itself in a higher issue alignment for multipliers. Furthermore, they amplify selected content for Twitter's recommendation algorithms, thus potentially pushing this content into feeds of users that do not even follow them. As \citet{Bouchaud2023} have shown, retweets play a significant role in the algorithmic amplification of content on Twitter/X, which may explain the high retweet activity of multipliers. While we may only speculate about the precise logic by which they act, we have found further evidence for the importance of multipliers in organizing political communication. At several occasions, influencers from both camps advise their followers to start following a given set of accounts, most of which are multipliers. This points to their importance in political campaigning on social media, and could be first evidence towards them being a part of of ``coordinated inauthentic behavior'' \citep{Cinelli2022}. More thorough analyses need to be conducted in order to investigate this type of behavior for multipliers in our dataset.\\
We nevertheless found evidence that point towards authenticity of the observed multiplier accounts. Following previous literature on bot-like behaviour on social media \cite{Elmas2022}, we analyzed account suspension, coordinated account creation, and circadian activity patterns. Only 6\% of multiplier accounts are suspended as of November 2024. Furthermore, we could not find evidence for coordinated account creation, as the maximum number of accounts created in one day is 3. Finally, only 4 multiplier accounts exhibit strongly suspicious circadian patterns. Even though these examinations cannot fully rule out inauthentic behavior, they lead us to assume that the majority of observed multiplier accounts are not (traditional) bots.\\
What remains, regardless of their authenticity, is the importance of multipliers in relaying and amplifying a precisely curated set of messages, thus directly affecting the online public sphere. This role of users acting as intermediaries on digital platforms \citep{friemel2023public} has not been properly studied empirically, but our work suggests that these users play a central role in opinion formation. They are neither public figures nor ``regular'' social media users. So far, only insulated works have investigated this phenomenon. Examples include concepts such as partisan ``curation bubbles'' of news content \citep{green2023curation} or the role and influence of super\textit{commenters} \citep{hsiao2022participatory}. The efforts of these users might aim to create ``new forms of incidental exposure [that] might emerge'' \cite[p. 299]{trilling2019conceptualizing}. They appear to arrange content in such a way that it fits their political views -- as we find, even more systematically than the opinion leaders they follow.\\[2mm]
The presented study has limitations that call for further research into polarization and issue alignment on social media. For one, we only observe \textit{structural} polarization and do not enter a textual analysis of the different stances expressed on a given issue. Complementing the structural retweet-network based measurement with text-based stance detection using natural language processing on the tweets could provide further validation of the opinion camps we discovered. Furthermore, we cannot explain the possible discursive reasons of the observed issue alignment. This question may be answered by a more thorough analysis of the tweets, for instance by analyzing how the different issues are connected argumentatively.\\
On a more technical level, there is an interesting distinction that this work has not made. As we saw it in the case of migration, some users from the left-leaning cluster were retweeted \textit{into} the right-wing cluster because the issue is owned by their camp. This is arguably different than landing in that cluster by retweeting right-wing content. The difference between active and passive sorting into one or the other cluster should be further explored.\\
Broadening the analysis to a wider user set, the user alignment of the random sample of users taking part in more than 10 trends shown in Fig.~\ref{fig:useralignment} hints towards the fact that the consistent polarization observed for influencers and multipliers also persists for ``regular users''. The main difficulty in studying regular users is the fact that they are typically present in a smaller number of topics than influencers, let alone multipliers. We therefore have not made any statements about issue alignment for those users, and we keep the thorough analysis of their alignment as a subject of further research. However, we provide first insights that exhibit similar patterns to those of power users in Fig.~\ref{fig:issuealignment-regular}. It shows the issue alignment of a subset of users that takes part in retweet networks of at least 10 different topics. We see the same general pattern as for the high-activity users: overall alignment strong alignment, going from a strong core of topics related to \textit{Journalism}, \textit{Covid} and \textit{Right-Wing Extremism} to less aligned topics related to \textit{Social Politics} and \textit{Ukraine}. While these findings are preliminary, they allow for the hypothesis that the strong signals of polarization and issue alignment hold for the majority of the user base analyzed in this work.\\
Finally, we may ask whether similar patterns of polarization driven by highly active users extend to other social media platforms. On Facebook, \citet{Hindman2022} show that there exist influential hyper-active users that do not correspond to traditional public figures or influencers. Those ``superusers'' are capable of impacting algorithmic amplification of content in similar ways as the multipliers on Twitter. In light of these apparent parallels, more empirical work is needed to address these phenomena, work that could inform platform decisions to reduce the potentially negative impact of power users.
\section*{Acknowledgements}
AP and EO acknowledge funding by the European Union’s Horizon Europe programme under grant agreement ID 101094752: Social Media for Democracy (SoMe4Dem) - Understanding the Causal Mechanisms of Digital Citizenship. AP acknowledges funding by the French government under management of Agence Nationale de la Recherche as part of the “Investissements d’avenir” program, reference ANR-19-P3IA-0001 (PRAIRIE 3IA Institute). FG acknowledges funding from the ‘Digital News Dynamics’ research group at Weizenbaum Institute for the Networked Society in Berlin, funded by the Federal Ministry of Education and Research. 
\section*{Data and Code Availability}
We share anonymized data and code that allows to reproduce the alignment results at \url{https://github.com/pournaki/twitter-trends-icwsm25}.
\section*{Author Contributions}
Data collection: AP, FG. Conceptualisation: AP, FG, EO. Data analysis: AP, EO. Writing of the manuscript: AP. Review of the manuscript: AP, FG, EO.
\bibliography{references.bib}
\section*{Broader Impact and Ethical Considerations}
\label{sec:ethics}
Analyzing the mechanisms and drivers of polarization on social media is a major scientific effort that this work takes part in. Thoroughly describing large-scale patterns of engagement leading to a polarized online public sphere can help both policy-makers better assess the dangers of social media for democracy, as well as inform regulatory actors for useful measures to counter potentially harmful behavior online.\\ 
The public tweets collected in this study were analyzed and described in this paper following the GDPR, since no personal information related to on any accounts were mentioned in the paper. The data were, as is most often the case in big data experiments, collected without the formal consent from the 700,025 users that contributed to it. Even though their Twitter activity is public, they may not formally agree to be part of the scientific analysis \citep{rogers2018social}. Especially considering the sensitive nature of the presented computations, like the clustering into ideological camps, which could be misused by nefarious actors for political profiling, we choose to systematically refrain from mentioning any kind of personal information that may lead to the identity of users.\\
\appendix
\section{Appendix}
\renewcommand{\thefigure}{A\arabic{figure}}
\setcounter{figure}{0}
\renewcommand{\thetable}{A\arabic{table}}
\setcounter{table}{0}
\begin{table}[ht]
\caption{Number of training documents, F1 scores of logistic classifier (LC) and stochastic gradient descent (SGD) for each topic.}
\begin{tabular}{lrrr}
\toprule
Topic & $N_{\mathrm{docs}}$ & F1 (LR) & F1 (SGD) \\
\midrule
Drug Legalisation & 565 & \textbf{0.97} & 0.96 \\
Abortion & 128 & 0.96 & \textbf{0.97} \\
Energy & 1031 & 0.95 & 0.95 \\
Sports & 6094 & \textbf{0.95} & 0.94 \\
Ukraine & 5723 & \textbf{0.95} & 0.94 \\
Migration & 610 & 0.94 & 0.94 \\
Covid & 4462 & \textbf{0.94} & 0.92 \\
Social Media & 1381 & 0.93 & 0.93 \\
Police & 887 & \textbf{0.93} & 0.92 \\
Climate Change & 2514 & 0.91 & \textbf{0.93} \\
Greetings and Holidays & 2187 & 0.90 & 0.90 \\
Social Politics & 1522 & \textbf{0.90} & 0.89 \\
Trans Lgbtq & 1940 & 0.90 & 0.90 \\
Racism & 961 & 0.90 & \textbf{0.93} \\
Religion & 904 & 0.89 & 0.89 \\
Gaming & 986 & 0.89 & \textbf{0.90} \\
Right Wing Extremism & 905 & 0.88 & 0.88 \\
Foreign Politics & 2760 & \textbf{0.87} & 0.84 \\
German Politics & 4503 & \textbf{0.87} & 0.86 \\
Music & 598 & 0.86 & 0.86 \\
Antisemitism & 609 & 0.86 & \textbf{0.89} \\
Journalism Media & 1025 & \textbf{0.85} & 0.83 \\
Outlier & 3820 & \textbf{0.83} & 0.82 \\
Pop Culture & 1385 & 0.81 & \textbf{0.82} \\
Democracy & 1339 & \textbf{0.79} & 0.78 \\
Mobility & 519 & 0.78 & \textbf{0.80} \\
\bottomrule
\end{tabular}
\end{table}
\subsection{(Semi-) supervised topic model}
\label{appendix:topicmodel}
In this section, we describe the topic model procedure in detail for reproducibility. We denote the full corpus of original tweets as $\mathcal{C}_{f}$. First, we extracted the top 50 most retweeted tweets from each trend that will be part of the representative corpus $\mathcal{C}_{r} \subset \mathcal{C}_{f}$. On this corpus, we inferred an unsupervised topic model using Python library BERTopic. After removing URLs and reply markers from the tweets, we encoded each tweet using the pre-trained multilingual sentence-embedding model `paraphrase-multilingual-MiniLM-L12-v2'. Then, we reduced the dimensionality of the sentence vectors to 5 dimensions using UMAP, using the following parameters: n\_neighbors=15, min\_dist=0.0, metric=cosine. We then run BERTopic with the following HDBScan parameters: min\_cluster\_size=100, min\_samples=1, metric=euclidean. This combination of sentence embedding model and parameters yielded 143 topics on this corpus. 44\% of the documents are labeled as outliers. This is due to the feature of HDBScan that identifies points that are in between several clusters as outliers. In our case, we left the tweets labeled as outliers aside in this first step of topic inference. We repeated these steps for the two other multilingual sentence-embedding models `paraphrase-multilingual-mpnet-base-v2' and `distiluse-base-multilingual-cased-v2' and chose `paraphrase-multilingual-MiniLM-L12-v2' due to the smallest number of outliers and inference speed. Using the cluster hierarchy provided by HDBScan as guidance, we merged the topics into larger categories, or issues, which resulted in the 25 issues described in Tab.\ref{tab:topicmodel}, and one topic that we considered as actual ``outliers'' (tweets that contain no topical information). This resulted in a set of 57,433 vector-label tuples that we could use for supervised topic modeling. We trained two different classifiers on these tuples and evaluated it using 10-fold cross-validation. The F1-scores for the logistic classifier and the stochastic gradient descent based classifier are reported in Tab.~\ref{tab:topicmodel}. Both performed well in predicting the topic of unseen documents, we chose the logistic classifier for its slighly higher mean F1 score (0.893 vs 0.891). With this classifier at hand, we encoded all original tweets from $\mathcal{C}_{f}$ using the same sentence-embedding model for $\mathcal{C}_{r}$, reduced their dimensionality to 5 using the same UMAP model, and inferred the topic labels. Finally, we assigned one topic label to each trend based on the distribution of topics in the tweets. These trend-topic pairs were manually checked at the very last step of the pipeline. Small misclassifications of single tweets do not play a significant role for the overall pipeline, since the central purpose of the topic model is to assign an issue to a whole trend. The final result after manual checking is shown in Tab.~\ref{tab:topicmodel}.
\subsection{Silhouette score}
\label{appendix:silhouette}
The silhouette of a node measures how close a node is to their own cluster, compared to how distant they are to the opposing cluster \citep{Rousseeuw1987}. We define it for the case of two clusters $A$ and $B$. Given a node $i \in A$, we define the mean distance of $i$ to all other nodes in $A$ as 
\begin{equation}
    a(i) = \frac{1}{ | A | -1}\sum_{i\in A, j \neq i}d(i,j)
\end{equation}
where $d(i,j)$ is the Euclidean distance between $i$ and $j$. Similarly, we define the mean distance of $i$ to all nodes in the opposing cluster $B$ as
\begin{equation}
    b(i) = \frac{1}{ | B |}\sum_{i\in B}d(i,j)
\end{equation}
The silhouette $s(i) \in [-1,1]$ is then defined as 
\begin{equation}
    s(i) = \frac{b(i)-a(i)}{\max \{ a(i),b(i)\}  }
\end{equation}
The silhouette score $S(C)$ of a given clustering $C$ is then given by
\begin{equation}
    S(C) = \frac{1}{N} \sum_{i=1}^N s(i)
\end{equation}
\subsection{Opinion extraction from retweet networks through force layout and stochastic block model}
\label{appendix:sbm}
For each trend in our dataset, we extracted the retweet network, where vertices are users and two vertices $i$ and $j$ are connected by a directed link if $i$ retweets $j$. We discarded networks with less than 50 users from our further analysis. This results in 1726 networks for which we have to assess whether they are best described by 1 or 2 clusters. As a first pre-processing step, we remove those nodes that only retweet one user and are never retweeted, as they do not contribute anything to the global structure of the network. We remember their sole neighbor and assign them the cluster of this neighbor. For each pre-processed network, we computed a two-dimensional force-directed layout embedding of each network using Matlab\footnote{The rest of our pipeline is written in Python. To the best of our knowledge, there is no reasonably fast implementation of Force Atlas 2 in Python.}. The force-directed layout tends to position densely connected groups of nodes close to each other, making the cluster structure of the network easily apparent. Previous work has connected force layouts to latent space models \citep{Gaisbauer2023}, which may allow for the interpretation of clusters in such layouts as groups of users sharing similar positions in an underlying latent political space. In order to systematically extract the clustering, we infer a degree-corrected stochastic block model (SBM) constrained on $N_{\mathrm{blocks}} \in {1,2}$ using the Python library \texttt{graph-tool}. We repeat the inference 10 times. Each time the SBM inference yields 2 clusters, we compute the silhouette score to validate the inferred clustering using the previously computed force-directed layout embedding. Finally, we choose the clustering $K$ with the highest silhouette score from the 10 runs. If the silhouette score $S(K) \leq 0.4$, then we keep the clustering. Else, we discard it and assume that the network is best described by only a single cluster. The threshold of 0.4 was found through manual exploration. In a final step, we manually examine each embedding-clustering pair by cycling through the network plots, coloring the nodes by their clustering. In 13 cases, the visual clustering was not successfully recovered. We re-ran the inference for those cases for a few more runs until the SBM yielded the correct clusters.
%
\subsection{Influencers and multipliers}
\label{appendix:infmul}
\begin{figure}[t]%
  \includegraphics[width=.45\textwidth]{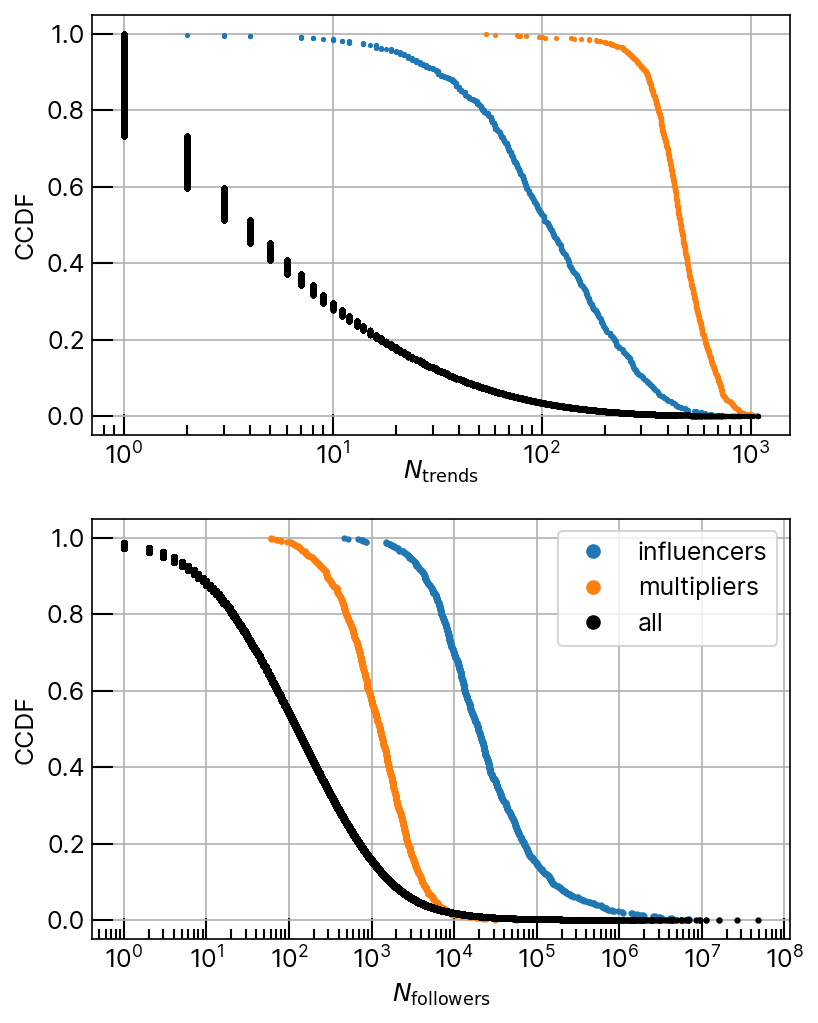}
  \caption{Complementary cumulative distribution (CCDF) of $N_{\mathrm{trends}}$ in which users participate (top plot) and $N_{\mathrm{followers}}$ (bottom plot), for all users (black), influencers (red) and multipliers (green). Multipliers participate in more trends than average users, influencers have more followers than average users.}
  \label{fig:ccdf}
\end{figure}%
\begin{figure}[t]%
  \includegraphics[width=.49\textwidth]{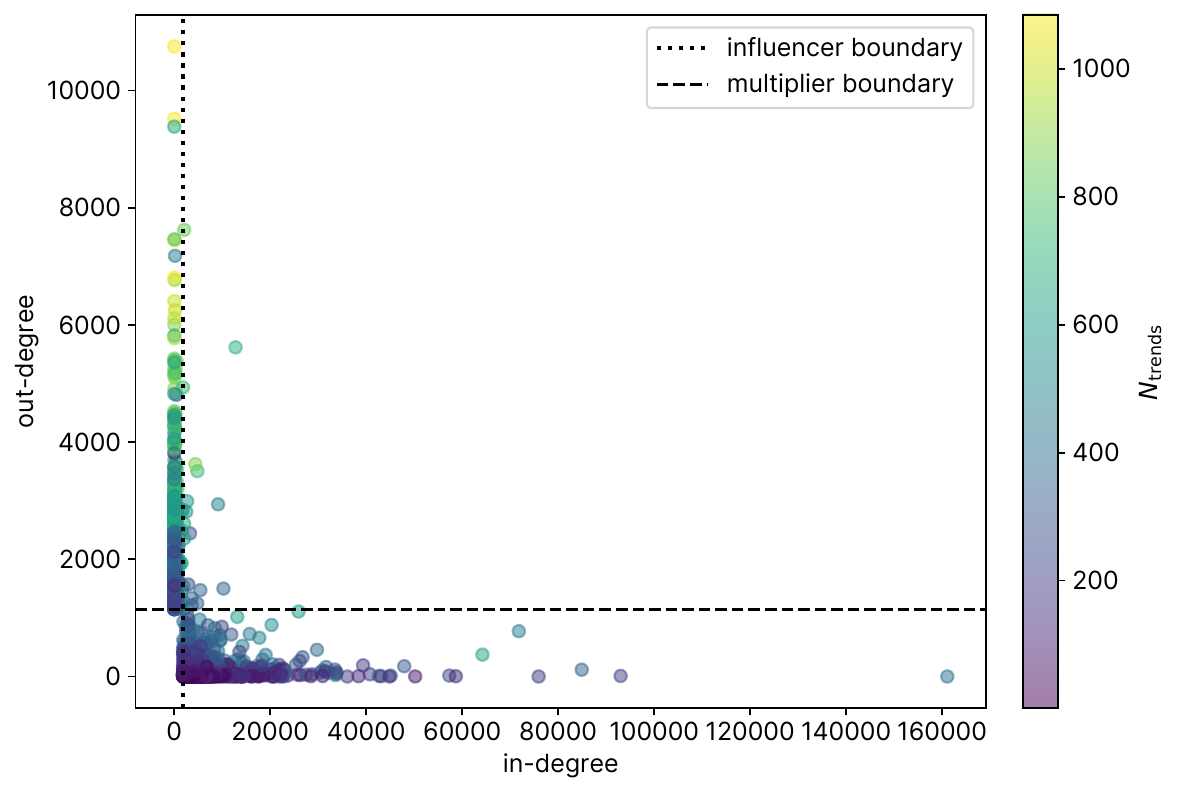}
  \caption{Overall out-degree as a function of overall in-degree of influencers/multipliers across trends. }
  \label{fig:appendix:infmul_overlap}
\end{figure}%
\subsubsection{Comparison to ``regular users''}
In order to compare the activity of influencers and multipliers to regular users, we plot the complementary cumulative distribution of $N_{\mathrm{trends}}$ and $N_{\mathrm{followers}}$. Multipliers are systematically present in a higher number of trends than influencers, who in turn are present in more trends than regular users. Furthermore, influencers have the most followers, but multipliers have significantly more followers than regular users.

\subsubsection{Overlap between influencers and multipliers}
One could argue that being an influencer or a multiplier is exhibition of certain behaviour and that users switch these roles from trend to trend. In our case, we measure this behaviour across the whole dataset. Figure~\ref{fig:appendix:infmul_overlap} shows the overall out-degree of users as a function of their overall in-degree. The boundaries for consideration as influencer (min. in-degree $ = $ 1833) and multiplier (min. out-degree $ = $ 1150) are shown as vertical/horizontal lines. We see that the diagonal is largely unpopulated, meaning that at the extremes, these two types of behaviour are almost mutually exclusive. There are only 19 users that exhibit both types of behaviour across the dataset. 

\subsection{Authenticity of influencers and multipliers}
While a thorough analysis of authenticity/bot-like behaviour of accounts corresponding to influencers and multipliers is outside the scope of this work, we provide some markers leading us to assume that the majority of accounts are authentic.
\subsubsection{Account suspension and creation}
First, we examine the number of accounts that are still active in both camps as of 2024-12-07, 15 months after the original trend data collection. More than 80\% of the influencer/multiplier accounts are still active, the share being a bit higher for influencers. There are more suspended multipliers than influencers. Previous work on ``bot farms'' showed that accounts from such organized farms tend to exhibit the same creation date \citep{Elmas2022}. The maximum number of same-leaning multiplier accounts created in one day is 3, leading us to assume that the observed multipliers do not belong to bot farms.
\subsubsection{Circadian patterns}
Next, we examine the circadian patterns of (re)tweet behaviour. We first compare the average number of tweets per half-hour of all users, influencers and multipliers. We see no stark differences between the distributions, except for multipliers being slightly more active at night. We then perform a manual analysis of individual users' circadian patterns, defining two criteria as strong markers of bot-like behaviour: constant activity throughout the day or systematic regularity of activity at specific times of the day (e.g. hourly posting). Using this method, we find only 4 multiplier accounts that are most certainly bots. We attach the circadian patterns as supplementary material.
\begin{figure}[t]%
  \includegraphics[width=.49\textwidth]{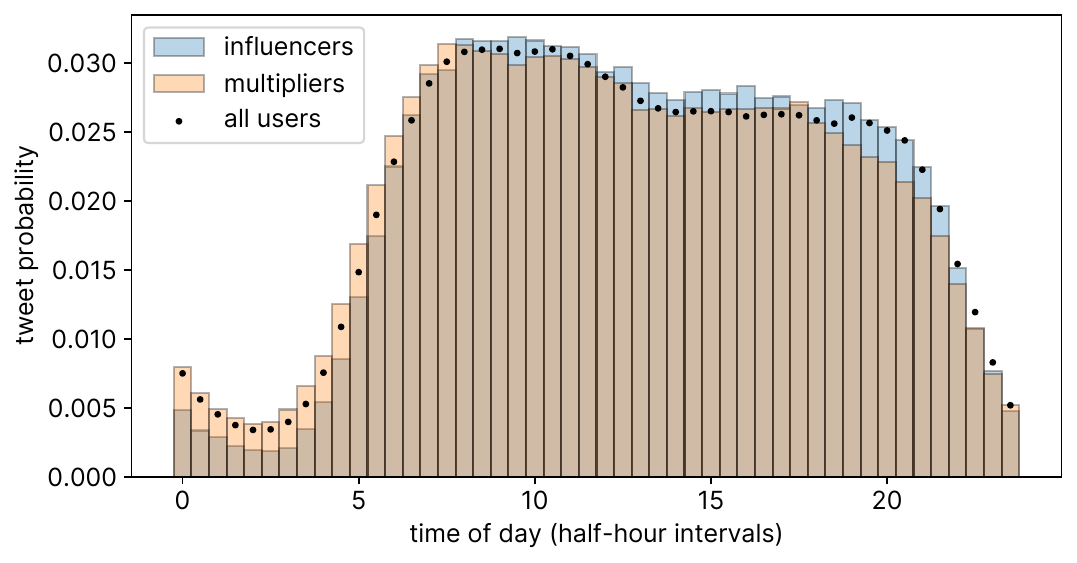}
  \caption{Average circadian pattern of influencers, multipliers and all users in the dataset.}
  \label{fig:circadian}
\end{figure}
\begin{figure*}[t]%
  \includegraphics[width=\textwidth]{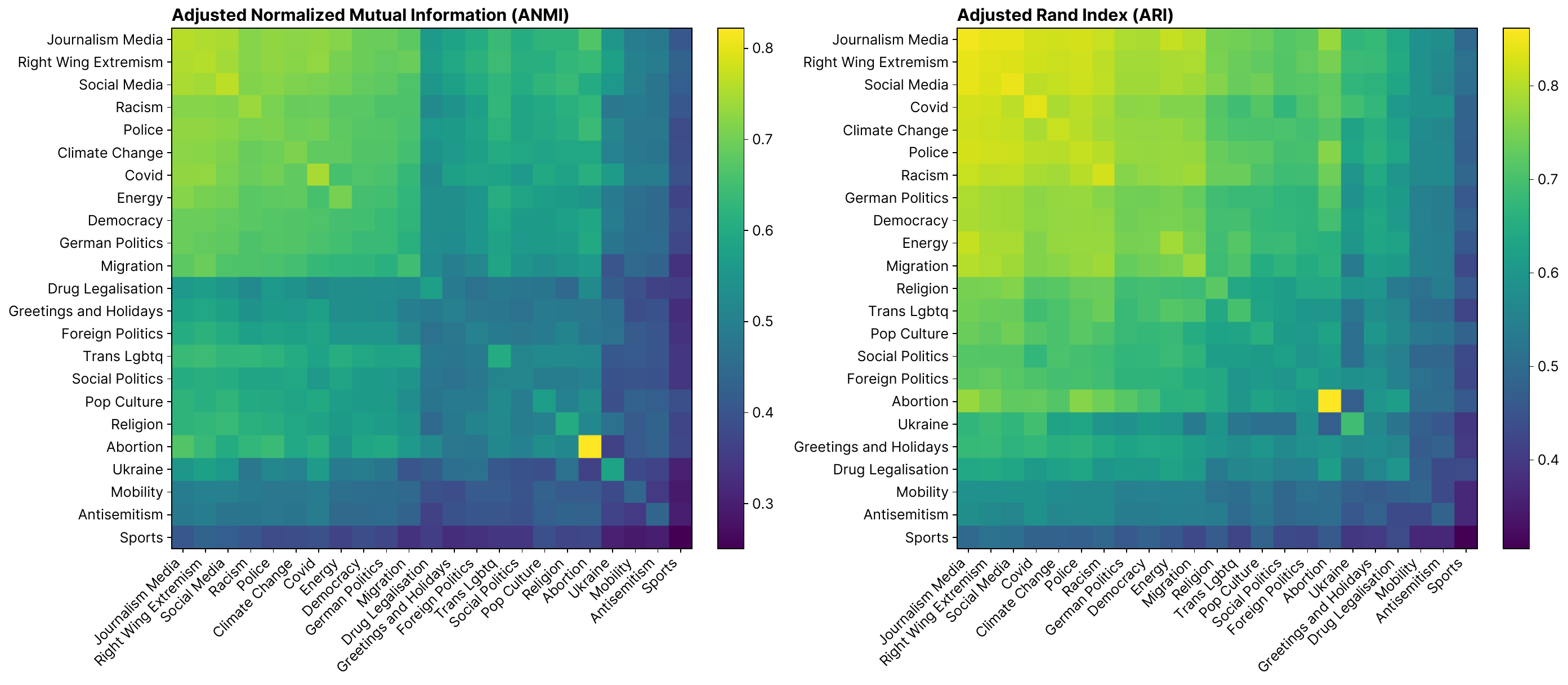}
  \caption{Issue alignment computed using the ANMI (left) and ARI (right). Abortion is very strongly internally aligned since there are only two polarized trends for this topic.}
  \label{fig:appendix:anmiari}
\end{figure*}%
\begin{table}[h]
      \caption{Number of active, deleted, suspended accounts for influencers and multipliers on 2024-12-06 (15 months after the original trend data collection). While the number of still active accounts is similar between influencers and multipliers, there are more suspended accounts for multipliers.}
\begin{tabular}{lrrr}
\toprule
 & \textbf{active} & \textbf{deleted} & \textbf{suspended} \\
\midrule
\textbf{influencers} & 880 & 87 & 33 \\
\textbf{multipliers} & 847 & 89 & 64 \\
\bottomrule
\end{tabular}
    \label{tab:my_label}
\end{table}
\subsection{Alternative approaches to computing issue alignment based on retweet networks}
\label{appendix:alignment_measures}
In the main text of the paper, we presented a user-centric approach to measure issue alignment on a set of topic-categorized graphs. This approach has several advantages compared to previously used methods by \citep{Chen2021,Salloum2024} based on graph partition similarity. For one, the user-based measure can be computed on unpolarized networks, which is useful for cases in which only one of the two camps is present in a trend. Furthermore, graph partition similarity measures can only be computed on the overlap of users between two trends, which can only represent a fraction of the graph.\\
In the following section, we present two alternative approaches to computing issue alignment, based on similarity measures between partitions. We derive one information-theoretic and one count-based measure, closely following previous work by \citep{Meila2003} and \citet{Vinh2010}. Computing these measures on our dataset shows qualitatively similar results to the method presented in the paper.

\subsubsection{Normalized Mutual Information (NMI)}
The following definition of the mutual information draws heavily on \citep{Meila2003}. Let us consider a partition (or clustering) $C$ of a set of points $D$ into $K$ subsets $S_1,...,S_K$, such that $S_i \cap S_j \neq \varnothing $. Let $n$ and $n_k$ be the number of points in $D$ and $K$ respectively. If we pick a point at random, the probability of this point being in cluster $k$ is equal to:
\begin{equation}
  \label{eq:a1}
  P(k)=\frac{n_k}{n}
\end{equation}
This random variable is uniquely associated to the clustering $C$. We can define the entropy of the random variable:
\begin{equation}
  \label{eq:a2}
  H(C)=-\sum_{k=1}^{K} P(k) \log P(k)
\end{equation}
It is always non-negative and becomes 0 when there is only one cluster. We can now define the mutual information between two partitions $C$ and $C'$, whose associated random variables we denote $P(k)$ and $P'(k')$. Let us define the probability of a point to be in cluster $k$ in $C$ and cluster $k'$ in $C'$:
\begin{equation}
  \label{eq:3}
  P\left(k, k^{\prime}\right)=\frac{\left|C_{k} \cap C_{k^{\prime}}^{\prime}\right|}{n}
\end{equation}
The mutual information $I(C,C')$ between to clusterings is equal to the mutual information of the the associated random variables:
\begin{equation}
  \label{eq:4}
  MI\left(C, C^{\prime}\right)=\sum_{k=1}^{K} \sum_{k^{\prime}=1}^{K^{\prime}} P\left(k, k^{\prime}\right) \log \frac{P\left(k, k^{\prime}\right)}{P(k) P^{\prime}\left(k^{\prime}\right)}
\end{equation}
It measures the reduction of uncertainty of $C'$ when we know $C$. It is non-negative and symmetric, and it can never exceed the entropy of any of the partitions:
\begin{equation}
  \label{eq:5}
  MI\left(C, C^{\prime}\right) \leq \min \left(H(C), H\left(C^{\prime}\right)\right)
\end{equation}
This score can be normalized by dividing it by the mean of the entropies 
\begin{equation}
    NMI\left(C, C^{\prime}\right) = \frac{2 I\left(C, C^{\prime}\right)}{H(C)+H(C^{\prime})}
\end{equation}
\subsubsection{Rand Index}
Another common approach to compare two partitions consists in counting pairs of nodes that appear in the same cluster in both partitions. Formally, let us define the number of node pairs that are in the same cluster in partition $C$ and $C^{\prime}$ as $N_{11}$. Conversely, $N_{00}$ denotes the number of node pairs that are in different clusters in both partitions. These two cases can be considered ``positive examples''. $N_{10}$ denotes the number of node pairs that are in the same cluster in $C$ but in different clusters in $C^{\prime}$, and $N_{01}$ the number of node pairs that are in a different cluster in $C$ but in the same cluster in $C^{\prime}$. These are the ``negative examples''. The Rand score is defined as 
\begin{equation}
    R(C,C^{\prime}) = \frac{N_{11}+N_{00}} {N_{00}+N_{11}+N_{10}+N_{01}}
\end{equation}
\subsubsection{Adjustment for chance}
Neither of the presented scores account for the fact that the clusters could be simply assigned randomly. Ideally, one would assign a score of 0 for random assignment of nodes to clusters. This can be achieved by adjusting the score for chance, which in turn implies the selection of an appropriate null model. One commonly used null model is the permutation model \citep{Lancaster1969}, where alternative clusterings are generated randomly based on the observed clustering while keeping the number of clusters and number of nodes in each cluster fixed. We refer the reader to \citet{Vinh2010} and \citet{Hubert1985} for a detailed derivation of the expected values for the Normalized Mutual Information and the Rand Index, leading to the Adjusted Normalized Mutual Information (ANMI) and Adjusted Rand Index (ARI). For the sake of general understanding, we provide the general formula for adjustment for chance as presented in \citet{Hubert1985}:
\begin{equation}
    \mathrm{adjusted~score} = \frac{\mathrm{score} - \mathrm{expected~score}}{\mathrm{max.~score} - \mathrm{expected~score}}
\end{equation}
\subsubsection{Computation}
In order to test the validity of the results obtained using our issue alignment measure, we compute the issue alignment using the Adjusted Normalized Mutual Information (ANMI) and the Adjusted Rand Index (ARI). For each index, we compute the pairwise index between all retweet networks and define the alignment between two issues as the average value of that index over all pairs of trends that belong to the issue. Both the ANMI and the ARI score are implemented in \texttt{sklearn}. The results are shown in Figs.~\ref{fig:appendix:anmiari}. They do not qualitatively differ from the results obtained using the user-centric measure. 
\subsection{Computation times and machines used}
\label{appendix:computationtimes}
Encoding the 4M original tweets from $\mathcal{C}_{f}$ using `paraphrase-multilingual-MiniLM-L12-v2' took 32m on a machine equipped with an NVIDIA A100 on an institute-internal cluster. Computing the force-directed layouts of the 1726 retweet networks took about 6h on a Thinkpad T14s (AMD Ryzen 7 PRO 5850U, 32GB RAM) using Matlab 2022a. The rest of the computations were done on an institute-internal server equipped with 4x 16-Core Intel Xeon E7-8867 v3 at 2.50 GHz. The SBM inference for the same networks took 1h30m. The computation of the alignment for regular users in Fig.~\ref{fig:issuealignment-regular} took 3h.
\subsection{Possible artifacts in the data used}
\label{appendix:artifacts}
We identify two sources of artifacts in our data collection process. First, the data collection server did not collect any tweets in August of 2022. Second, since the collection was conducted using a single API key, tweets of trending topics were only collected for 24 hours every day. It is therefore possible that, for certain days that generated trends with unusually high numbers of tweets, we were not able not collect all 5 identified top trends, as the collection was stopped after 24 hours. However, we do not believe either of these two aspects affects the overall findings presented in the paper.
\end{document}